\documentclass[aps,prl,showpacs,twocolumn,superscriptaddress,nolongbibliography,floatfix]{revtex4}
\usepackage{graphicx}
\usepackage{amsmath}
\usepackage{xparse}
\usepackage{dcolumn}
\usepackage{bm}
\usepackage[colorlinks=true, citecolor=blue,allcolors=blue]{hyperref}
\usepackage{physics}

\newcommand{\ii}{\mathrm{i}}

\DeclareUnicodeCharacter{251C}{-}
\DeclareUnicodeCharacter{2310}{-}
\DeclareUnicodeCharacter{0393}{-}
\DeclareUnicodeCharacter{2560}{-}
\DeclareUnicodeCharacter{255D}{-}

\begin{document}

\title{
A Continuous Pump-Probe Experiment to Observe Zeeman Wave Packet Dynamics
} 

\author{K.\,L.~Romans}
\affiliation{Physics Department and LAMOR, Missouri University of Science \& Technology, Rolla, MO 65409, USA}

\author{K.~Foster}
\affiliation{Physics Department and LAMOR, Missouri University of Science \& Technology, Rolla, MO 65409, USA}

\author{S.~Majumdar}
\affiliation{Physics Department and LAMOR, Missouri University of Science \& Technology, Rolla, MO 65409, USA}

\author{B.\,P.~Acharya}
\affiliation{Physics Department and LAMOR, Missouri University of Science \& Technology, Rolla, MO 65409, USA}

\author{O.~Russ}
\affiliation{Physics Department and LAMOR, Missouri University of Science \& Technology, Rolla, MO 65409, USA}

\author{A.\,H.\,N.\,C.~De Silva}
\affiliation{Physics Department and LAMOR, Missouri University of Science \& Technology, Rolla, MO 65409, USA}

\author{D.~Fischer}
\affiliation{Physics Department and LAMOR, Missouri University of Science \& Technology, Rolla, MO 65409, USA}

\date{\today}

\begin{abstract}
In this work, we study the coherent dynamics of an atomic Zeeman wave packet using a continuous pump--probe scheme. A polarized wave packet is generated via few-photon excitation by a femtosecond laser pulse, creating a state with a magnetic moment tilted relative to an external magnetic field. The subsequent Larmor precession of the atoms is probed by continuous ionization in the field of an optical dipole trap (ODT) laser. Photoelectrons and photoions are detected in coincidence using a cold target recoil ion momentum spectrometer (COLTRIMS). While the addition of the ODT enables further cooling of the atomic ensemble, it removes the pulsed timing reference typically used to extract photoelectron momentum distributions in COLTRIMS. Here, we present a method that extends the standard COLTRIMS technique by exploiting redundancy in the measured data to reconstruct the time of ionization. The resulting time-dependent ionization signal reflects the coherent precession of the atomic magnetic moment, enabling real-time access to atomic dynamics on nanosecond timescales.
\end{abstract}


\maketitle

Time-resolved laser spectroscopy, along with high-resolution momentum imaging, provides detailed insights into the dynamics of atomic or molecular systems. Femtosecond pump-probe schemes were pioneered by A. Zewail \cite{Zewail2000} and they evolved into the gold standard for time-resolved spectroscopy. They were extensively used to study nuclear dynamics in molecular reactions on a femtosecond timescale (e.g. \cite{Stolow2004}). Shorter time scales became accessible with the availability of attosecond laser pulses created by high-harmonic generation \cite{Corkum2007, Kling2008, Krausz2009} using streaking \cite{Itatani2002, Cattaneo2016}, pump-probe \cite{Tzallas2011, Geneaux2019, Duris2019}, or the interferometric RABBITT \cite{Paul2001, Muller2002, Cattaneo2016, Cattaneo2022} techniques. Nowadays, there is a wealth of time-resolved methods exploiting different types of light sources ranging from accelerator-sized free-electron lasers, to compact tabletop laser devices employing advanced pulse shaping techniques, and spanning time-scales from the attosecond to the picosecond time regime.

Recently, we reported an alternative experimental approach for time-resolved photoelectron spectroscopy \cite{Romans2023}, operating on the comparatively slow nanosecond timescale and thus providing a complementary perspective to the ultrafast techniques discussed above. Unlike conventional pump-probe schemes, where time information is obtained by scanning the delay between two pulses, our method employs a femtosecond pump pulse in combination with a weak continuous-wave (cw) probe. The cw probe can photoionize the excited atomic system at any time after the pump pulse, enabling time-resolved measurements without the need for delay scanning.

To detect photoelectrons and recoiling target ions in coincidence, we employ a cold target recoil ion momentum spectrometer (COLTRIMS) \cite{Ullrich2003, Moshammer2003, Fischer2012, Fischer2019}. In conventional COLTRIMS, time-of-flight information is required to reconstruct the momentum vectors of the fragments. In our approach, however, we rely solely on the measured hit positions of the particles on the detectors. A dedicated reconstruction algorithm allows us to recover not only the energy and angular distributions of the photoelectrons, but also the time of ionization relative to the femtosecond pump pulse, thereby providing additional insight into the system's time evolution.

While our previous work \cite{Romans2023} focused on the incoherent population dynamics driven by spontaneous decay, we now investigate the coherent dynamics of the wave packet formed by the pump pulse. By using an energetically narrow laser pulse for excitation of a Zeeman wavepacket, we are able to track the electron's motion following the excitation. Such systems are of key importance for the development and study of coherent control schemes in ultracold atomic samples (e.g., \cite{Takei2016, Devolder2021}).

\begin{figure}[!t]
\centering
\includegraphics[keepaspectratio, width=0.95\linewidth]{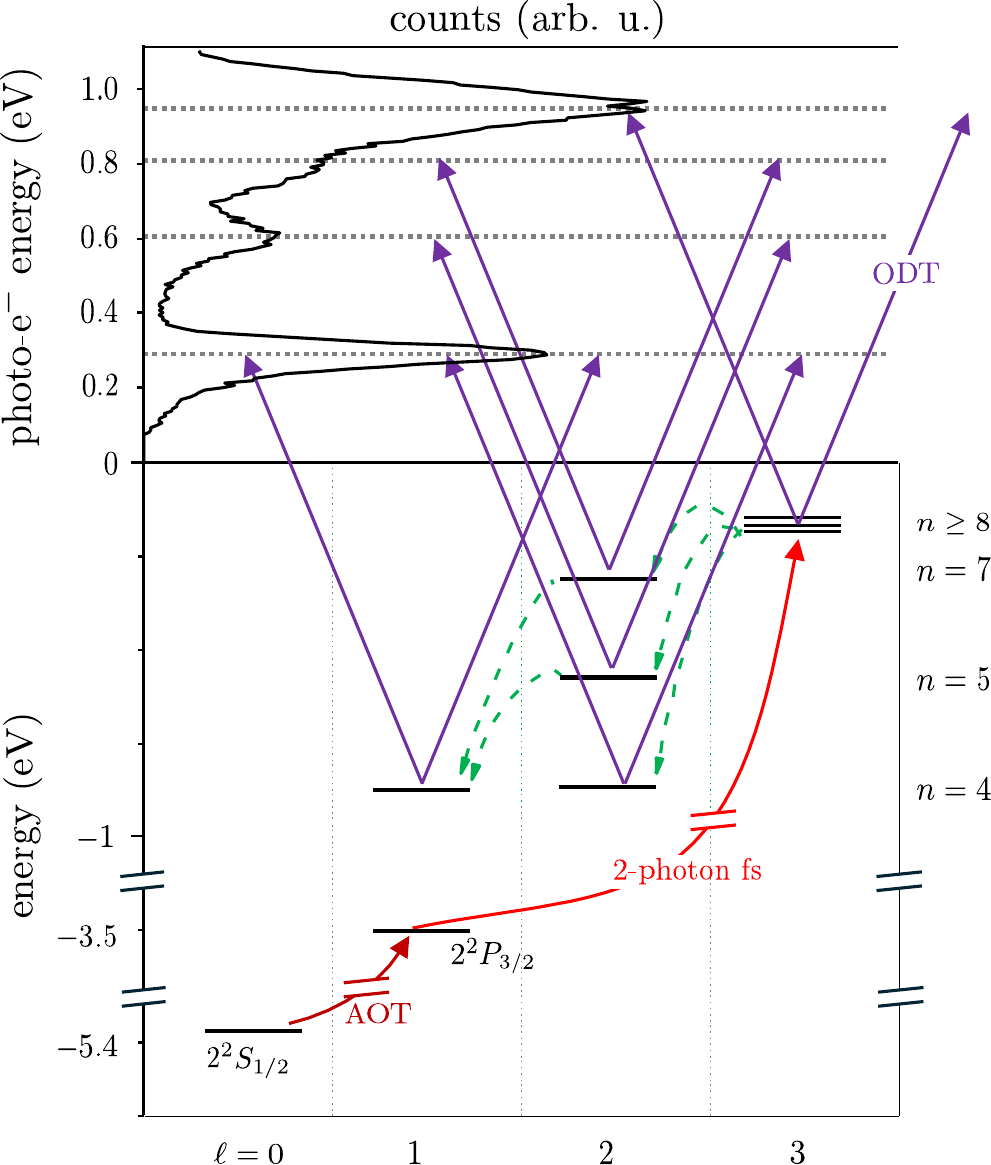} 
\caption{Ionization pathway (bottom) and corresponding photoelectron energy spectrum (top). Continuous-wave AOT and ODT photons are shown in maroon and purple, respectively, while the femtosecond pulse-induced transition is depicted in red. Dominant spontaneous decay channels are indicated by dashed arrows.}
\label{fig:energy levels}
\end{figure}

Specifically, we measure the time evolution of magnetic Zeeman wave packets in a weak, homogeneous external magnetic field. The general pathway is depicted in Fig.~\ref{fig:energy levels}. First, atomic $^6$Li gas is suspended in the laser light of a near-resonant all-optical trap (AOT); for details of the trap, see \cite{Sharma2018}. Within the AOT, the atoms are optically pumped, resulting in a population of about 25,\% in the excited $2^2P_{3/2}(F=5/2)$ hyperfine state. Of these excited atoms, more than 93,\% occupy a single orbital magnetic sublevel, $m_{\ell} = -1$, defined with respect to the quantization axis---the $z$-axis---which is given by the direction of the magnetic field.

The excited Zeeman wave packet is generated by two-photon absorption from a femtosecond light source, which transfers population to an excited $f$-state with principal quantum number $n \geq 8$. The femtosecond laser system is similar to that described in \cite{Harth2017} and is based on a commercially available optical parametric chirped-pulse amplifier (OPCPA). At its heart is a Ti:Sa oscillator that provides a broadband seed signal for two noncollinear optical parametric amplifier (NOPA) stages. For this study, the light source was tuned to emit pulses with a central wavelength of $735 \pm 10$~nm, a pulse duration of $50$~fs, and a repetition rate of $200$~kHz. Unlike in previous experiments (e.g., \cite{Silva2021, Acharya2021}), the intensity of the source was kept below the threshold for direct few-photon ionization, with a peak intensity of up to $10^{11}$~W/cm$^2$. The beam intersects the AOT at an angle of $12.5^{\circ}$ with respect to the $z$-axis.

Finally, the atoms are photoionized in the continuous-wave field of an optical dipole trap (ODT) laser, either directly from the highly excited state or after spontaneous decay to a lower-lying state. The ODT is generated using an industrial-grade ytterbium-doped fiber laser, which outputs an infrared continuous wave at $1070 \pm 5$~nm. The laser beam's polarization and propagation directions lie in the $xy$-plane, and it is focused into the reaction volume to a beam waist of approximately 50,$\mu$m, resulting in an intensity of about $10^{7}$~W/cm$^2$.

The momenta of photoelectrons and photoions are measured using a cold target recoil ion momentum spectrometer. Standard COLTRIMS relies critically on time-of-flight (TOF) information for recoil ions and electrons, denoted $T_{r/e}$, which is typically measured relative to a pulsed projectile (e.g., a laser) source. In the present experiment, however, photoionization occurs in the continuous-wave field of the optical dipole trap (ODT) laser, making direct TOF measurement infeasible. Nevertheless, TOF information can be recovered by exploiting the fact that, due to the low temperature of the atomic cloud and the negligible momentum of the ionizing photon, the total momentum of each ion--electron pair can be assumed to be zero \cite{Hubele2015}. As a result, the full six-dimensional momentum space of the photoelectron and photoion reduces to three independent dimensions, and their measured hit positions on the respective detectors are sufficient to reconstruct the complete kinematics of the fragmentation process.


The details of the time-reconstruction method are described elsewhere \cite{Romans2023, Romans2023a}; here, we provide only a brief summary. In the spectrometer, charged fragments are directed toward opposing particle detectors by homogeneous electric and magnetic fields aligned along the longitudinal $z$-axis. Due to the Lorentz force, particles with charge $q$ undergo cyclotron motion in the magnetic field, which has a single non-zero component, $B_z$. As a result, the $x$ and $y$ positions of particles with initial transverse momentum components $p_x$ and $p_y$ after a time-of-flight $T$ are given by \cite{Fischer2019}:
\begin{align}
    x &= \frac{1}{ q B_{z} }\left[ \sin(\omega_{c} T)\,p_{x} + (1-\cos(\omega_{c} T)) p_{y} \right], \label{eq:x}\\
    y &= \frac{1}{ q B_{z} }\left[ (\cos(\omega_{c} T)+1) p_{x} +  \sin(\omega_{c} T) p_{y} \right].\label{eq:y}
\end{align}
Here, $\omega_{c}=qB_z/m$ is the cyclotron frequency of the particles. Their motion along the $z$-direction is governed by the electric field, which has its only non-zero component $E_z$ along this axis:
\begin{align}
    z=\frac{p_z}{m}T+\frac{qE_z}{2m}T^2.\label{eq:z}
\end{align}
In this equation, $z$ represents the distance between the reaction volume and the particle detector.

The key to the present time-reconstruction method lies in the fact that, although the momenta of the ion and electron are equal in magnitude (due to momentum conservation), and despite their motion being governed by the same set of equations (Eqs.~(\ref{eq:x})--(\ref{eq:z})), their trajectories in the spectrometer's electromagnetic field are highly asymmetric. This asymmetry arises from the significantly greater mass of the recoil ion compared to that of the photoelectron. 

This becomes most apparent when considering the cyclotron phase angle from Eqs.~(\ref{eq:x}) and (\ref{eq:y})
\begin{equation}
\omega_{c} T = \frac{B_z}{E_z\sqrt{m}} \left( \sqrt{ \frac{p_z^2}{m} + 2zqE_z } - \frac{p_z}{\sqrt{m}} \right),
\end{equation}
which is obtained by solving Eq.~(\ref{eq:z}) for $T$.
As can be seen from the expression on the right-hand side within the parentheses, the phase angle's dependence on the initial momentum $p_z$ becomes weaker for heavier particles.

In fact, the ions undergo only a fraction of a cyclotron revolution, with their phase angle varying by less than 1,\textperthousand\ of $2\pi$. In contrast, the much lighter electrons undergo several full cyclotron revolutions before reaching the detector, and their final rotational phases $\omega_c T$ depend strongly on their initial momenta. Essentially, the electrons' rotational phase---and hence their time of flight---is encoded in the relative positions of the recoil ions and electrons on their respective detectors. If the variation in phase remains within a single 2$\pi$ cycle, the electrons' travel time can be unambiguously recovered, allowing for the reconstruction of their full three-dimensional momentum vectors.

\begin{figure}[!htbp]
\centering
\includegraphics[keepaspectratio, width=\linewidth]{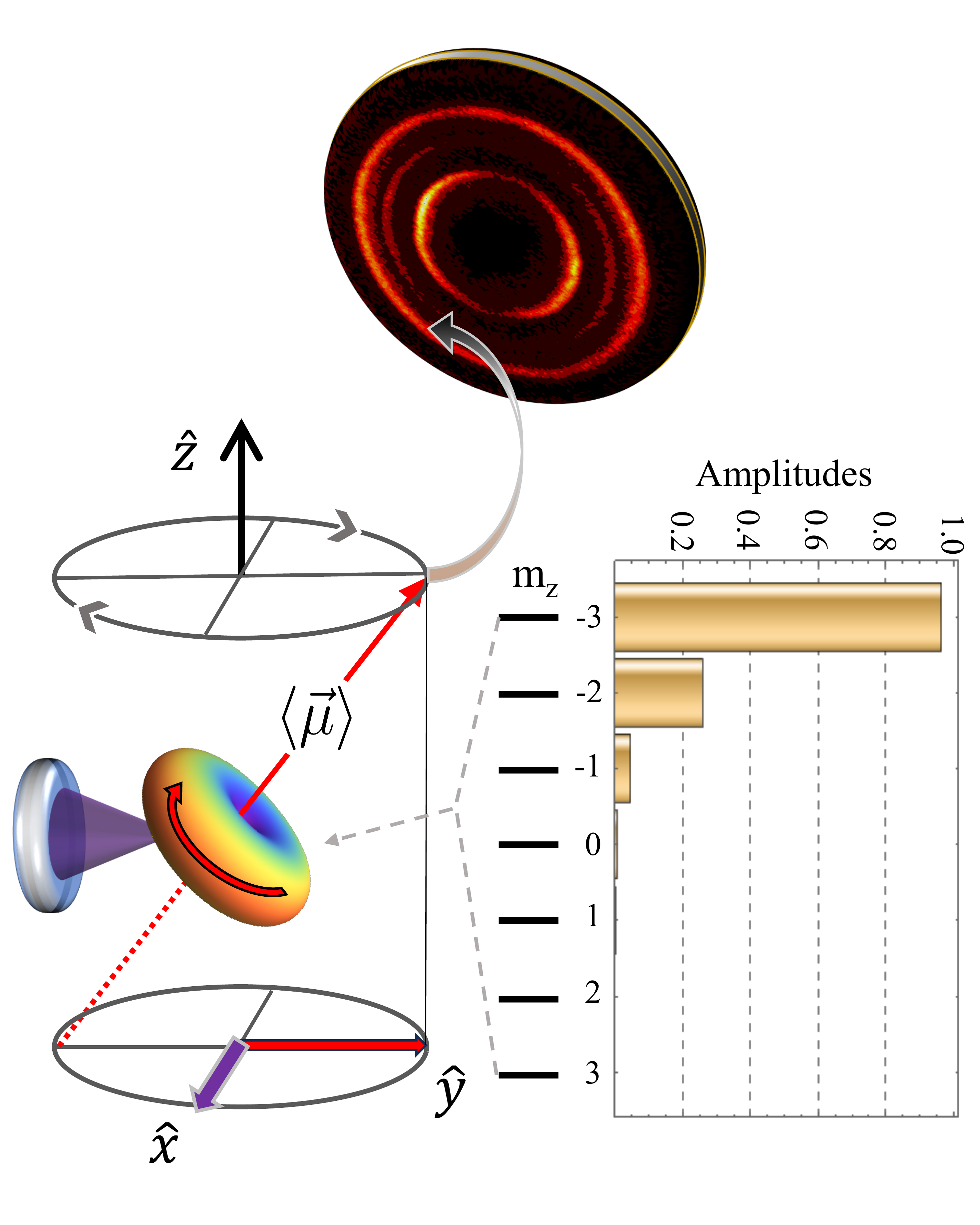} 
\caption{Qualitative depiction of the rotating magnetic moment induced by the femtosecond excitation (left), the initial population of magnetic sublevels (right), and the corresponding recovered photoelectron transverse momentum distribution (top).}
\label{fig:rotating moment}
\end{figure}

A systematic test of this method is reported in \cite{Romans2023, Romans2023a}, where an excellent agreement (within 5\,ns) was observed between directly measured and reconstructed times of flight. On the one hand, this timing information can be used to compute the resulting energy and momentum distributions (see the top panels of Fig.\ref{fig:energy levels} and Fig.\ref{fig:rotating moment}, respectively). The different photoelectron energies reflect the population distribution of the excited system following femtosecond excitation and, potentially, subsequent spontaneous decay. On the other hand, the recovered time information opens access to a broader class of time-dependent phenomena. One example is the slow, microsecond-scale population dynamics driven by spontaneous decay cascades. By focusing on the dominant decay channels, it becomes straightforward to extract explicit inter-state population dynamics in an effective real-time scheme \cite{Romans2023}.

\begin{figure}[!htbp]
\centering
\includegraphics[keepaspectratio, width=\linewidth]{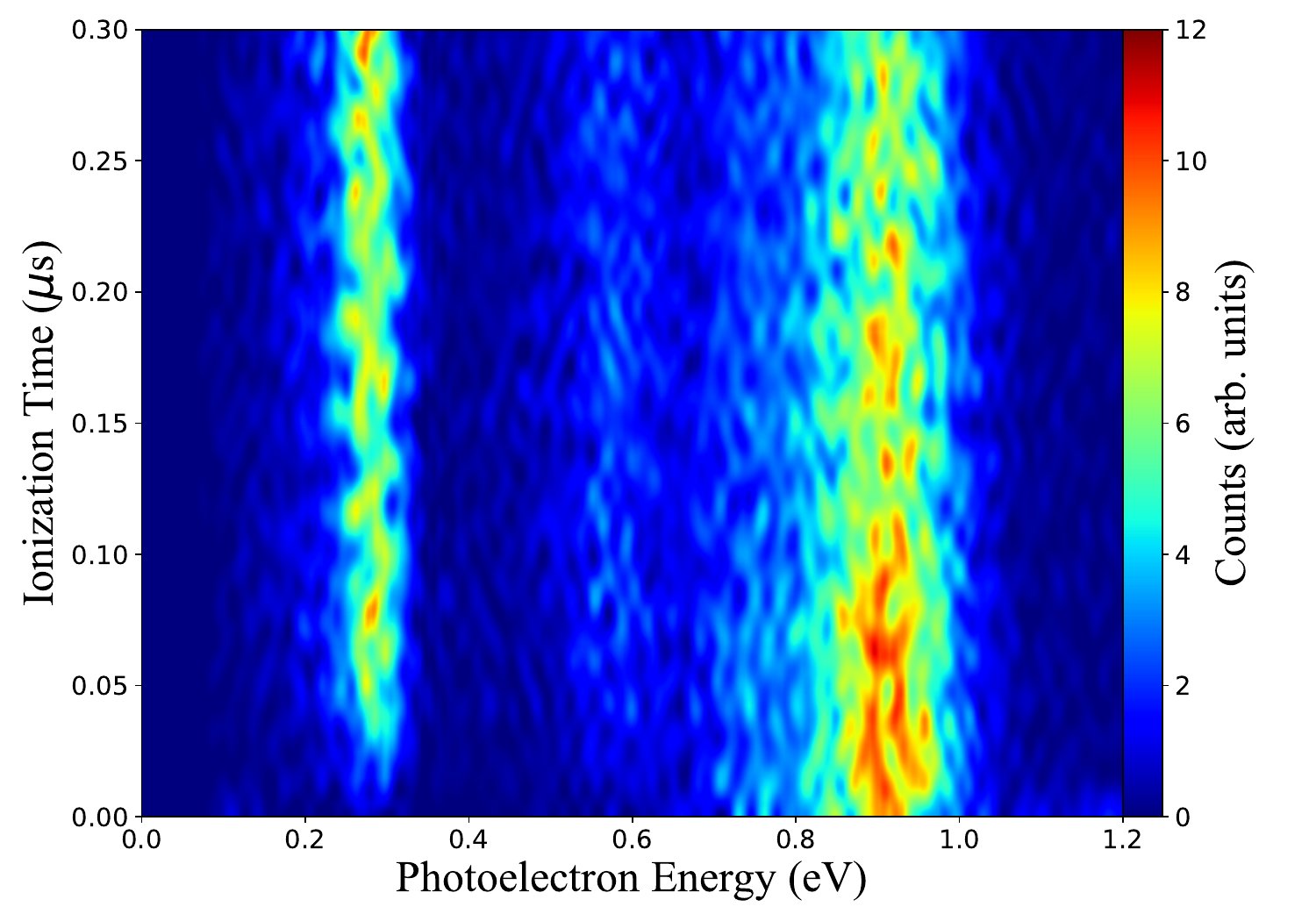} 
\caption{Ionization time delay vs.\ photoelectron energy, with the $z$-axis representing the rate of electron emission in the $xy$-plane.}
\label{fig:iontime vs energy}
\end{figure}

The nanosecond accuracy also allows for the exploration of faster atomic dynamics ($\sim 1$ ns), which can be seen in Fig.~\ref{fig:iontime vs energy}. Here, the cross section of electrons being emitted in the $xy$-plane is plotted as a function of photoelectron energy and the derived time of ionization. The ionization rate features a periodic time dependence throughout the whole energy range. Two dominant streaks appear, which can be associated with the $n=4$ states on the left (centered at about 0.3\,eV) and the $n\geq 8$ superposition on the right (at about 0.9\,eV). Comparing with Fig.~\ref{fig:energy levels}, the $5$D and $7$D states can be seen at around $0.6$ and $0.8$ eV, respectively; the remaining weaker lines come from the copious intermediate states in the cascade and are not shown along with the dominant decay channels. 

This experimental data can be explained with the simple model shown in Fig.~\ref{fig:rotating moment}. The atoms are exposed to the homogeneous magnetic spectrometer field oriented along the $z$-axis. The circularly polarized femtosecond laser beam is guided through the chamber in the $yz$-plane exciting the atoms to a excited wavepacket. This momentarily maintains the alignment of the atoms' mean magnetic moment $\expval{\Vec{\mu}}$ along the photons' propagation direction. Since this moment has a component perpendicular to the magnetic field the system will undergo Larmor precession around the $z$-axis. Meanwhile, the atoms have sufficient energy to absorb a single infrared photon from the ODT field and be ionized. The infrared photon propagates along the $y$-axis, with its linear polarization along the $x$, shown in purple. The probability of ionizing the atom and emitting photoelectrons in the $xy$-plane depends on the relative orientation of this photon's polarization and the atom's magnetic moment. Given the geometry, two maxima and two minima are expected per Larmor precession cycle, corresponding to those times when the mean magnetic moment is aligned or perpendicular to the polarization axis, respectively.   

Formally, the excited wavepacket is initially prepared in an aligned $f$-states, i.e., $|m_{\ell}| = 3$ with respect to the quantization direction defined by the propagation direction of the femtosecond laser. This alignment is tilted by 12.5$^{\circ}$ relative to the laboratory $z$-axis. As a result, the wavepacket can be expressed as a coherent superposition of magnetic sublevels with $|m_{\textrm{z}}|\leq 3$, where $m_z$ refers to the orbital angular momentum's projection along the $z$-axis. This decomposition is illustrated in Fig.~\ref{fig:rotating moment}. The resulting state can be written as the following sum:
\begin{equation}
    \ket{\psi_{\mathrm{ex.}}(t)} = \sum_{n \geq 8}^{n_{\mathrm{max}}}\sum_{m_{\textrm{z}} = -3}^{3} c_{m_{\textrm{z}}}^{(n)}\,\exp(-\ii \,\Omega_{m_{\textrm{z}}}^{(n)}\,t)\ket{n,3,m_{\textrm{z}}},
    \label{eq:init state mix}
\end{equation}
where the $c_{m_{\textrm{z}}}^{(n)}$ and $\Omega_{m_{\textrm{z}}}^{(n)}$ are the expansion amplitudes and eigenenergy angular frequencies, respectively, while the $\ket{n,3,m_{\textrm{z}}}$ are the bound lithium $f$-orbitals. 

The ionization rate from this state into a final continuum state $\ket{\psi_{\varepsilon, \,\ell_f}}$ can be calculated using the equation of \cite{Starace2006},
\begin{equation}
    R_{\mathrm{ex.}}(t) \propto \left| \matrixel{\psi_{\varepsilon, \, \ell_f}}{D_{\lambda}}{\psi_{\mathrm{ex.}}(t)} \right|^2,
    \label{eq:ion rate calc}
\end{equation}
where $D_{\lambda}(\equiv\boldsymbol{\epsilon}\vdot\boldsymbol{r})$ is the electric dipole operator, and ($\varepsilon, \ell_f$) is the final state energy and angular momentum. When Eqs. \eqref{eq:init state mix} and \eqref{eq:ion rate calc} are combined, and the matrix element is integrated over the solid angle of the ejected photoelectron, the dominant time-dependent term becomes, 
\begin{equation}
    R_{\mathrm{ex.}}(t) \propto 1-A\cos\Bigl(2\cdot\left(\Omega_{- 3} - \Omega_{- 1}\right)t\, \Bigl) + \dots,
\end{equation}
with $A$ being a relative amplitude. In other words, the energy difference between the $m_z = -3$ and $-1$ magnetic sublevels drive the time variation of the ionization rate, with the next leading term, which depends on $m_z = -2$ and $0$, being two orders of magnitude smaller. 

In order to apply this model and assess its accuracy, the energy level shifts $\Omega_{m_{\textrm{z}}}^{(n)}$ must be calculated for different magnetic sublevels. Naively, one might expect that the $m_{\textrm{z}}$-dependence of these shifts arises solely from the Zeeman effect, which contributes a term $m_{\textrm{z}} \cdot \omega_L$, where $\omega_L \propto B_z$ is the Larmor frequency. However, this simplistic assumption underestimates the observed frequency, and for the present magnetic field strength of approximately 4\,Gauss, the calculated shift is a factor of two smaller than what is extracted from experimental data. 

However, besides coupling the atoms to the continuum via ionization, the ODT field can also couple lower-lying bound states and shift the energy levels via the dynamic Stark effect. These energy shifts can be approximated as a sum over two-state terms, $\Delta E_{DS}^{(n)} \approx \sum_{n'} \Delta E_{n, n'}$, for each of the dipole-allowed transitions between the $n'D$-states and $nF$-Rydberg levels. In the large detuning limit this coupling is given as \cite{Haas2006, Hertel2015},
\begin{equation}
   \Delta E_{n, n'} = -\frac{I_0}{2\epsilon_0 \hbar c} \frac{\left| \matrixel{\psi_{nF}}{D_{\lambda}}{\psi_{n'D}}
   \right|^2}{\delta_{n,n'}},
\end{equation}
where $\delta_{n,n'}$ is detuning of the laser frequency with repsect to the atomic transition, and $I_0$ is the ODT's intensity. Including the Zeeman and Autler-Townes effects in the calculated energy shifts leads to the predicted frequency picking up the missing factor of two and more closely mirroring the fits.
\begin{figure}[!t]
\centering
\includegraphics[keepaspectratio, width=\linewidth]{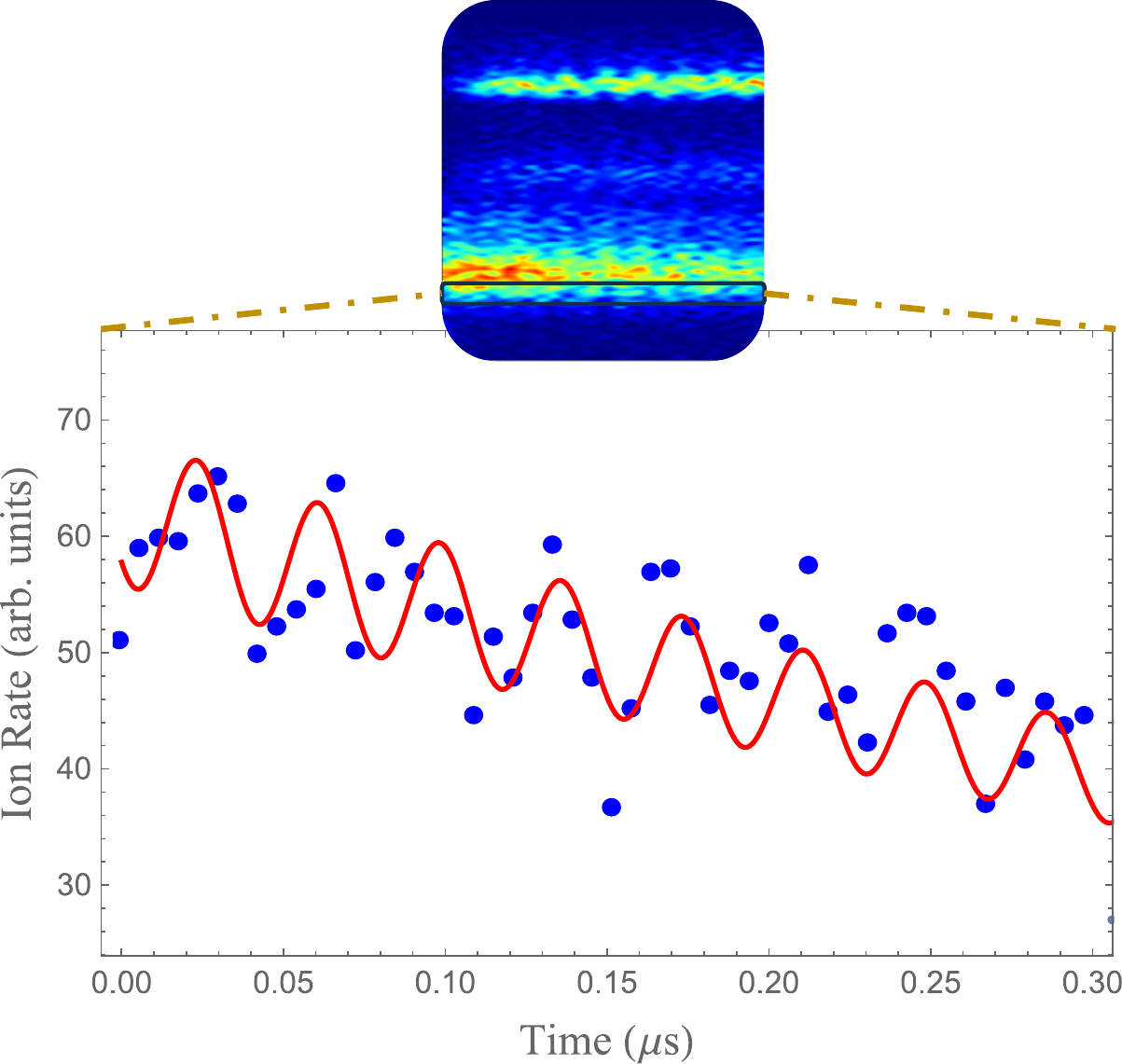} 
\caption{Time evolution of the ionization rate from states with principal quantum number $n \geq 8$. Experimental data points are obtained by slicing the data shown in Fig.~\ref{fig:iontime vs energy}, as indicated by the inset at the top. The prediction from a simplified single-state model is shown as a solid red line.}
\label{fig:dynamics}
\end{figure}
The results of the calculation can be seen in Fig.~\ref{fig:dynamics}, where an energy slice, corresponding to photoelectron energies of the highly-excited state, was explored. Assuming a single 8$F$ state, the model captures the oscillation frequency of $\approx 27$ MHz. The overall intensity drops over time due to spontaneous decay. 

With the success of this simple model, it begs the question as to the source of the periodicity in the remaining energy levels. While the states with $n\geq 8$ are excited coherently by the femtosecond laser, the lower-lying levels are populated predominately by incoherent spontaneous decay. Such a process should quickly destroy the phase information contained in the Rydberg superposition as it gets dispersed amongst the cascade. Despite this, the $n=4$ states near the $0.3$ eV peak retain a strong periodicity and seem to hold coherence throughout the experimental run. 

This is explained based on relative timescales. The directly highly-excited states populated in the fs-pulse are long-lived ($\sim 1$/,$\mu$s) as compared to the lower levels ionization rates in the ODT field. The $n=5,7$ states have rates roughly $10$-fold higher, while the $n=4$ states are $10$-fold higher still. This leads to the middle states having a time dependence that is a mix of the decay channels and the driving ODT field, while the short-lived states are ionized almost as soon as they are populated. Consequently, any time-dependence seen in the $n=4$ ionization data must be a delayed, but direct, projection of that seen in the Rydberg levels. The phase shift seen on the left and right sides of the $n=4$ data can also be explained by the $4P$ and $4D$ channels seen in Fig.~\ref{fig:energy levels}; the former goes through a delayed path with the $5$ and $7$D states, while the latter has a direct dipole-allowed path from the $n\geq 8$ states. 

Initially intuitively pleasing, the current model is not without shortcomings. First, the spectrometer's resolution limits which states can be distinguished. The closely spaced Rydberg states are difficult to disentangle and lead to challenges when developing the analytic model. Second, only those states in the dominant decay channels (see Fig.~\ref{fig:energy levels}) are considered for the Autler-Townes shift. Bound state spontaneous decay constants were calculated using the Alkali Rydberg Calculator (ARC) \cite{Sibalic2017} at each step of the cascade, and weak couplings were neglected. Finally, although the coupling to the lower-lying bound states was included for the Autler-Townes effect, the continuum coupling was not. Sometimes referred to as the electron's ponderomotive, or quiver, energy, this effect further shifts the energy levels due to an intense laser field. Presently, the low intensity of the ODT field allows for this shift to be safely ignored, but it may need more careful consideration in future studies.   

In conclusion, we have demonstrated a method of derived time that allows for the investigation of time-dependent atomic processes in real time. This technique utilizes the continuous field of an infrared ODT laser, in conjunction with a pulsed femtosecond pump, to constantly probe the energy states of the atoms. Lithium-6 atoms initially in a spin-polarized $2^2P_{3/2}$ state are excited by the fs-pump into similarly aligned $f$-states. This excitation forms a coherent superposition of Rydberg states, which then couples to the ODT's and spectrometer's fields. This results in shifts in the atomic energy levels via the Zeeman and Autler-Townes effects. Furthermore, the axis of alignment is slightly titled with respect to the spectrometer's $z$-axis. This adds to the mix a superposition of magnetic sublevels or a magnetic wave packet. Each wave packet, and its associated mean magnetic moment, is dressed by the ODT's field and undergoes an enhanced Larmor precession. In this way, the ionization rate becomes explicitly time-dependent, and through its analysis, both the relatively slow population and faster wave packet dynamics become unlocked. 

The ionization scheme presented here is suited to study the dynamics of many atomic Alkali systems. Particular interest has been placed on Rydberg atoms given their recent return into the proverbial limelight as vehicles for information technology. Along the way, great effort has been taken to focus on dipole-allowed configurations, due to the ease of the framework, but a tremendous amount of physics lies just beyond the forbidden. Recent work into orbital angular momentum (OAM) light beams (e.g. \cite{Fang2022}) have demonstrated a convenient source of photons with angular quantum number $\ell \geq 2$. When used with atomic targets, one must take great pains to keep the atoms on the OAM beam's axis of symmetry, such that the angular momentum transfer remains well-defined; atoms have such a small spatial extent, relative to the OAM beam, that they experience an effective plane wave otherwise. In contrast, Rydberg atoms can easily have wave functions that extend out to a few hundred atomic units. This leaves Rydberg atoms as prime targets to couple to OAM fields for dipole-forbidden investigations. This work, in conjunction with \cite{Romans2023, Romans2023a}, not only demonstrates an accessible source of Alkali Rydberg atoms for many ongoing COLTRIMS experiments but also a way in which to probe their dynamics continuously. 
\begin{acknowledgments} 
The experimental material presented here is based on work supported by the National Science Foundation
under Grants \hbox{No.~PHY-2207854} and \hbox{No.~PHY-1554776}.
\end{acknowledgments} 

\bibliography{TDIon} 

\end{document}